\documentclass[prl,aps,twocolumn,amsmath,amssymb]{revtex4}

\pdfoutput=1
\usepackage{hyperref}
\usepackage{graphicx}

\usepackage[usenames]{color}

\def\cal#1{\mathcal{#1}}
\def\eqq#1{Eq.~(\ref{#1})}

\def\f#1{Fig.~\ref{#1}}

\def\ed{\epsilon_{\rm d}}
\def\es{\epsilon_{\rm s}}

\def\eed{{\rm e}^{\beta \epsilon_{\rm d}}}
\def\ees{{\rm e}^{\beta \epsilon_{\rm s}}}

\def\emed{{\rm e}^{-\beta \epsilon_{\rm d}}}
\def\emes{{\rm e}^{-\beta \epsilon_{\rm s}}}

\def\e{{\rm e}}

\def\beq{\begin{equation}}
\def\eeq{\end{equation}}
\def\bea{\begin{eqnarray}}
\def\eea{\end{eqnarray}}

\begin{document}

\title{Self-assembly of multicomponent structures in and out of equilibrium}

\author{Stephen Whitelam$^1$\footnote{\tt{swhitelam@lbl.gov}}, Rebecca Schulman$^2$\footnote{\tt{rschulm3@jhu.edu}}, Lester Hedges$^1$}
\affiliation{$^1$Molecular Foundry, Lawrence Berkeley National Laboratory, 1 Cyclotron Road, Berkeley, CA 94720, USA\\
$^2$Chemical and Biomolecular Engineering, Johns Hopkins University, 3400 N Charles St, Baltimore, MD 21218, USA}

\begin{abstract}
Theories of phase change and self-assembly often invoke the idea of a `quasiequilibrium', a regime in which the nonequilibrium association of building blocks results nonetheless in a structure whose properties are determined solely by an underlying free energy landscape. Here we study a prototypical example of multicomponent self-assembly, a one-dimensional fiber grown from red and blue blocks. If the equilibrium structure possesses compositional correlations different from those characteristic of random mixing, then it cannot be generated without error at any finite growth rate: there is no quasiequilibrium regime. However, by exploiting dynamic scaling, structures characteristic of equilibrium at one point in phase space can be generated, without error, arbitrarily far from equilibrium. Our results thus suggest a `nonperturbative' strategy for multicomponent self-assembly in which the target structure is, by design, not the equilibrium one.
\end{abstract}

\maketitle

Many theories of phase change and self-assembly place at their heart the idea that dynamical trajectories follow low-lying paths on the free energy landscape connecting reactants and
products~\cite{becker1935kinetische,volmer1926keimbildung,binder1976statistical,zlotnick1994build,endres2002model,PhysRevLett.99.076102,hansen2006theory,evans1979nature,lutsko2010recent,shen1996bcc,lutsko2006ted}. This idea underpins rate equation theories~\cite{becker1935kinetische,binder1976statistical,zlotnick1994build,endres2002model}, classical nucleation theory~\cite{becker1935kinetische,volmer1926keimbildung} and density functional theory~\cite{hansen2006theory,evans1979nature,lutsko2010recent,shen1996bcc,lutsko2006ted} (which assume that a structure's morphology is determined by minima or low-lying paths on the underlying free energy landscape), and the conjecture of Stranski and Totomanow~\cite{stranski1933rate}, which states that a system, confronted by a set of free energy barriers, will evolve by crossing the lowest of them. These formal statements reflect the intuition that one can generate structures characteristic of equilibrium using a sufficiently `mild' nonequilibrium protocol. Many one-component systems indeed assemble in quasiequilibrium~\cite{wolde1999homogeneous,shen1996bcc,sear2007nucleation,wolde1997epc,lutsko2006ted} if they are not deeply supercooled~\cite{hagan2011mechanisms,grant2012quantifying} or plagued by slow particle dynamics~\cite{hedges2011limit}. However, a substantial literature suggests that {\em multicomponent} self-assembly is susceptible to kinetic factors even under conditions of weak driving~\cite{kremer1978multi,stauffer1976kinetic,PhysRevB.27.7372,schmelzer2004nucleation,schmelzer2000reconciling,scarlett2010computational,kim2008probing,scarlett2011mechanistic}. For instance, the Stranski-Totomanow assumption breaks down in a particular case of simulated binary colloid nucleation, where sluggish inter-species mixing prevents nuclei from establishing compositional equilibrium~\cite{sanz2007evidence,peters2009competing}. Also, binary crystals have been observed in simulation and experiment to grow out of compositional equilibrium, even under conditions `mild' enough to produce morphologically ordered structures~\cite{kim2008probing,scarlett2010computational,scarlett2011mechanistic}. To describe such assembly theoretically, one must account for dynamical processes that drive a system away from low-lying paths on the free energy landscape~\cite{kremer1978multi,stauffer1976kinetic,PhysRevB.27.7372,schmelzer2004nucleation,schmelzer2000reconciling,Fichthorn1991mc,peters2009competing,lutsko2011communication,lutsko2012dynamical}. On a practical level, one can ask under what conditions can precisely-defined multicomponent structures be self-assembled, if evolution even near a phase boundary leads to a structure not characteristic of the equilibrium one?

Here we study this question within a prototypical example of multicomponent self-assembly. We apply simulation and quantitatively accurate analytic theory to the fluctuating growth of a model lattice-based fiber built from red and blue blocks. We show that when compositional correlations of the equilibrium structure are not equal to those of the randomly-mixed material incident on the fiber, the former cannot be generated at a finite rate of growth. Moreover, structures that assemble even close to the phase boundary can be very different to equilibrium ones. This absence of a quasiequilibrium regime occurs despite the fact that fibers near the phase boundary grow in a `quasireversible' manner, displaying many unbinding events. However, by exploiting dynamic scaling connecting equilibrium and nonequilibrium parameter manifolds, defined structures can be generated without error arbitrarily far from equilibrium. The failure of a widely-made assumption for perhaps the simplest example of compositionally inhomogeneous self-assembly confirms the need for the development of dynamical theories~\cite{marconi1999dynamic,fraaije1993dynamic,lutsko2011communication,peters2011coupling,corezzi2009connecting,schmelzer2000reconciling,scarlett2011mechanistic} in order to describe the self-assembly of multicomponent structures, and challenges the idea that the equilibrium structure is the natural target for the self-assembly of components of many types. 

{\em Model.} We consider a 1-dimensional stochastic growth process
in which a fiber is built from red and blue blocks (\f{fig1}(a)). We
add blocks to the right-hand end of the fiber with rate
$c$ (concentration). Added blocks are blue with probability $p_{\rm
  blue}$, and red otherwise. We allow the rightmost block to detach
from the fiber with rate $\e^{-\beta \epsilon_i}$, which depends on
the nature of the rightmost bond of the fiber. Nearest-neighboring
blocks of the same color interact with energy $-\epsilon_i=-\es $, while the red-blue interaction is $-\ed$ (we set $\beta =1$ throughout). We implemented this stochastic process in simulations by removing the
terminal block with probability $p_{\rm remove} = 1/\left( 1+ c \,
  \e^{\beta \epsilon_i} \right)$, where $i={\rm s}$ or d as
appropriate, and otherwise adding a block to the fiber end. This process is not designed to model any particular physical system, but is instead a simple fluctuating protocol that captures the key kinetic constraint imposed upon a 3d structure in 3d space: material can be removed only from the interface between a structure and its environment~\cite{scarlett2011mechanistic}. By considering a lattice model in which complications of morphology and the possibility of internal block rearrangements are suppressed, we can explore directly how this constraint affects pattern generation.
\begin{figure}[t!]
\includegraphics[width=\linewidth]{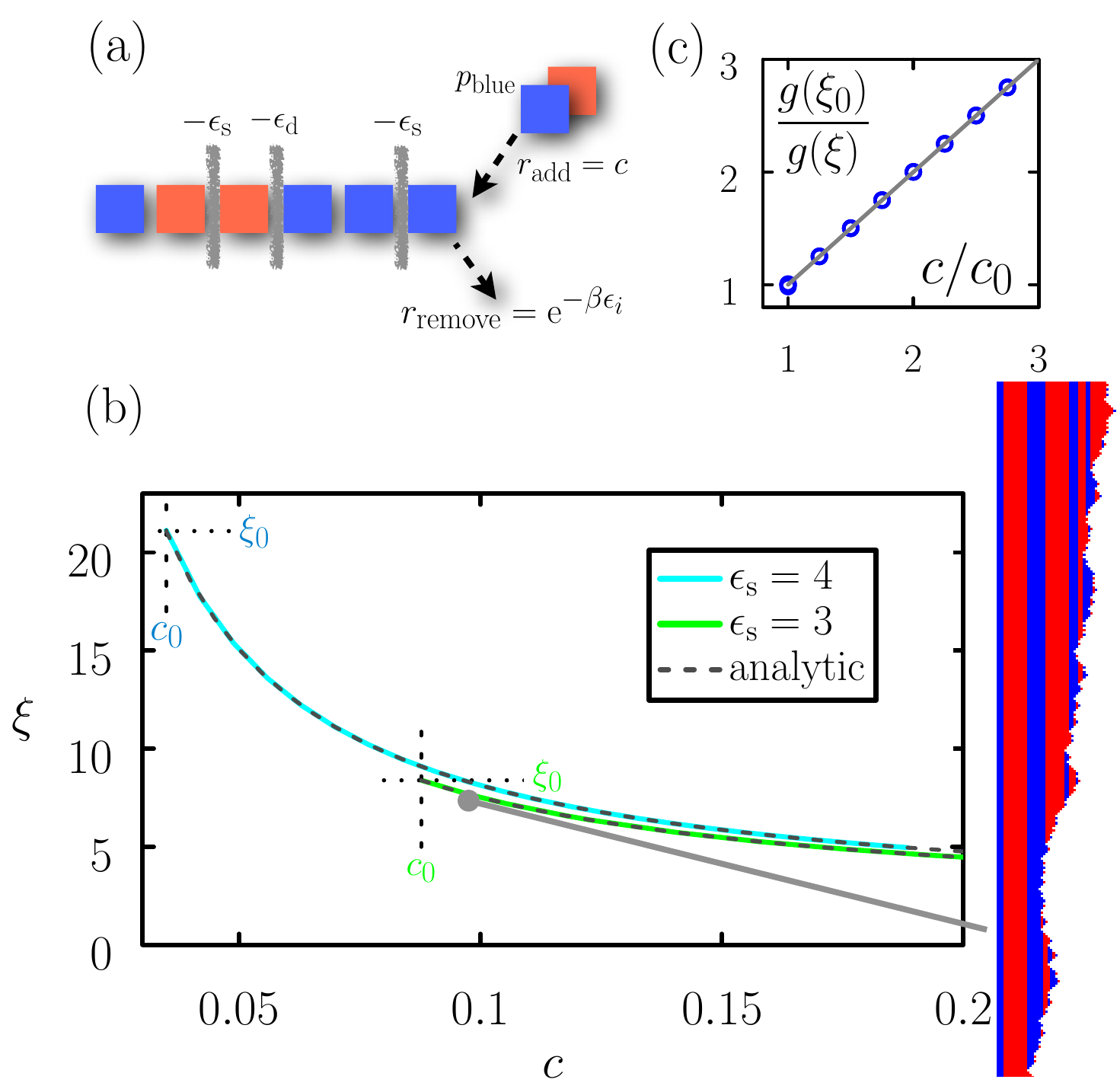}
\caption{\label{fig1} {\em Fiber structures depend sensitively upon preparation conditions.} (a) Schematic of fiber energetics and growth rates. (b) Domain lengths $\xi$ of dynamically-generated fibers of given energy scales $\es$ (simulations: solid lines; theory: dashed lines) are less than the corresponding equilibrium domain lengths $\xi_0$ (horizontal dashed lines) for all points past the phase boundary $c=c_0$ (vertical dashed lines). This breakdown of the quasiequilibrium assumption reflects the conflict between compositional correlations of unassembled material and the equilibrium structure, and occurs despite the fact that fibers close to the phase boundary grow in a `quasireversible' way. (c) Simulation data can be collapsed onto a master curve, summarizing the continuous variation of the scaled domain length $g(\xi) \equiv \xi (\xi-2)/(\xi-1)$ with concentration $c$.}
\end{figure} 

The `equilibrium' we consider is the one often imagined in theories of self-assembly, corresponding to minimization of the free energy of a structure of fixed size~\cite{wolde1999homogeneous,shen1996bcc,wolde1997epc,lutsko2006ted}. Such an equilibrium is achieved by our dynamical protocol in the limit of a large number of binding and unbinding events per site~\footnote{This protocol is unlike crystal growth done using, say, Metropolis Monte Carlo dynamics~\cite{scarlett2011mechanistic}, because the latter obeys detailed balance with respect to an underlying Hamiltonian. However, it is similar in the sense that both protocols achieve equilibrium only in the limit of zero growth rate.}: a key question is how the structure generated at a finite rate of growth compares to the equilibrium one. The energetics of a fiber of fixed size is that of the 1d Ising model~\cite{binney1992theory,chandler}, with Hamiltonian ${\cal H} = -J \sum_i S_i S_{i+1} - h \sum_i S_i$. Here the spin variable $S_i = \pm 1$ describes a blue $(S_i=1)$ or red $(S_i=-1)$ block; the coupling $J = (\es-\ed)/2$ is the penalty for domain wall (red-blue bond) creation; and the magnetic field $h = - \ln(1/p_{\rm blue}-1)/2$ describes the bias for blue blocks over red ones. Here we add red and blue blocks with equal likelihood, i.e. $p_{\rm blue} =1/2$, equivalent to $h=0$. In this case the equilibrium structure of a fiber consists of equal proportions of red and blue blocks arranged into domains whose lengths $\ell$ occur with probability $\rho_0(\ell) = (\xi_0-1)^{-1} \exp \left[ \ell \ln (1-\xi_0^{-1})\right]$. Here $\xi_0= 1+ \exp(2 \beta J)$ is the mean domain length in equilibrium.

{\em Results.} We carried out dynamic simulations for the choice $\ed=1$ and a range of values of $\es>\ed$ (for which $\xi_0 > 2$). For each set of energetic parameters we considered a range of concentrations $c$. When $c<c_0 = 2/(\ees+\eed)$ the fiber does not grow. When $c=c_0$ the drift velocity of the fiber, averaged over distances greater than a typical domain length, is exactly zero. The fiber therefore `grows' only by diffusion of its rightmost end. When $c>c_0$, the fiber grows with nonzero drift velocity. We therefore consider the concentration $c_0$ to define the `phase boundary' between non-assembly and assembly. We stopped dynamic simulations when a fiber of length $L=2.5 \times 10^4$ blocks was generated. We performed $10^4$ simulations for each concentration considered, except at the phase boundary, where the diffusive growth of a fiber was slow; there, we generated about 200 fibers for each set of conditions. For each ensemble of fibers we measured the probability distribution of domain lengths $\rho(\ell)=n(\ell)/\sum_{\ell=1}^L n(\ell)$, where $n(\ell)$ is the number of occurrences of domain length $\ell$ across all simulations at given thermodynamic conditions. These distributions were always exponential (\href{http://nanotheory.lbl.gov/people/fiber_paper/fiber_supp.pdf}{Fig. S1}) with a mean $\xi$ depending on all three parameters $c$, $\es$ and $\ed$. This mean is shown in \f{fig1}(b) for two choices of $\es$. At the phase boundary the equilibrium structure is generated. For all points past the phase boundary,  the dynamic domain length $\xi$ is less than the equilibrium one $\xi_0$, despite the fact that fiber growth close to the phase boundary is highly fluctional, exhibiting many unbinding events (\f{fig1}(b), snapshot at right). Moreover, fiber structures are exquisitely sensitive to preparation conditions, and change continuously with supersaturation.

Analytic theory reveals that this sensitivity is an inevitable consequence of the different compositional statistics of the equilibrium structure and unassembled material. Consider defect variables $\eta_i \equiv S_i S_{i+1}$, where $\eta_i=1$ describes a bulk (same-color) bond and $\eta_i=-1$ is a defect (unlike-color) one. Let $\phi \equiv [2\eta_i-1]$ be the likelihood that a given bond is a bulk one, where the average $[\cdot]$ is taken over many realizations of the dynamics. Enumeration of basic microscopic processes implies a drift velocity for bulk domains of
\beq
\label{eq1}
v_{\rm bulk} = c/2 - \phi \emes,
\eeq
and a drift velocity for the fiber
\beq
\label{eq2}
v_{\rm fiber} = c - \phi \emes-(1-\phi)  \emed.
\eeq

At the phase boundary, where $v_{\rm bulk} = v_{\rm fiber} =0$, Eqns.~(\ref{eq1}) and~(\ref{eq2}) give the equilibrium bulk fraction $\phi_0 = 1/(1+{\rm e}^{\beta( \epsilon_{\rm d} - \epsilon_{\rm s})})$, and the equilibrium concentration $c_0 =2/(\eed+\ees)$. $\xi_0$ and $c_0$ are given by horizontal and vertical dotted lines on \f{fig1}(b). Fiber {\em dynamics} can be solved by requiring $v_{\rm bulk}/v_{\rm fiber}=\phi$, yielding
\beq
\label{eq3}
\phi= \frac{c/2 - \phi \emes}{c - \phi \emes-(1-\phi)  \emed}.
\eeq
This equation is straightforwardly solved (\href{http://nanotheory.lbl.gov/people/fiber_paper/fiber_supp.pdf}{Eq. S1}) for the domain length $\xi = 1/(1-\phi$), and we plot this solution as grey dashed lines in \f{fig1}(b). The agreement with simulation is good, confirming the sensitivity of fiber structure to method of preparation. Moreover, the structure of \eqq{eq3} reveals the origin of this sensitivity. Its large-$c$ limit returns the domain length characteristic of random mixing, $\xi_\infty=2$, which is in
general unequal to the equilibrium domain length $\xi_0$. At the phase
boundary, the balance of $c$-dependent and -independent terms in
\eqq{eq3} is such that bulk domains are generated at a
rate characteristic of equilibrium. For finite supersaturation,
however, the statistics of the fiber begins to reflect the statistics
of random mixing:  \eqq{eq3} can be expanded in small deviations $\delta c \equiv c -c_0$ from the phase boundary to yield
\beq
\label{eq_pert}
 \xi-\xi_0 \approx - \frac{\xi_0}{c_0} \frac{(\xi_0-2)(\xi_0-1)}{\xi_0(\xi_0-2)+2} \delta c. 
  \eeq 
Thus fiber structure is a continuous function of concentration, and the dynamic correlation length is less than the equilibrium one for {\em any} concentration $c>c_0$. Equivalently, structures generated at finite growth rate always sit above the minimum of the free energy landscape (\href{http://nanotheory.lbl.gov/people/fiber_paper/fiber_supp.pdf}{Fig. S2}).

\begin{figure}
\includegraphics[width=\linewidth]{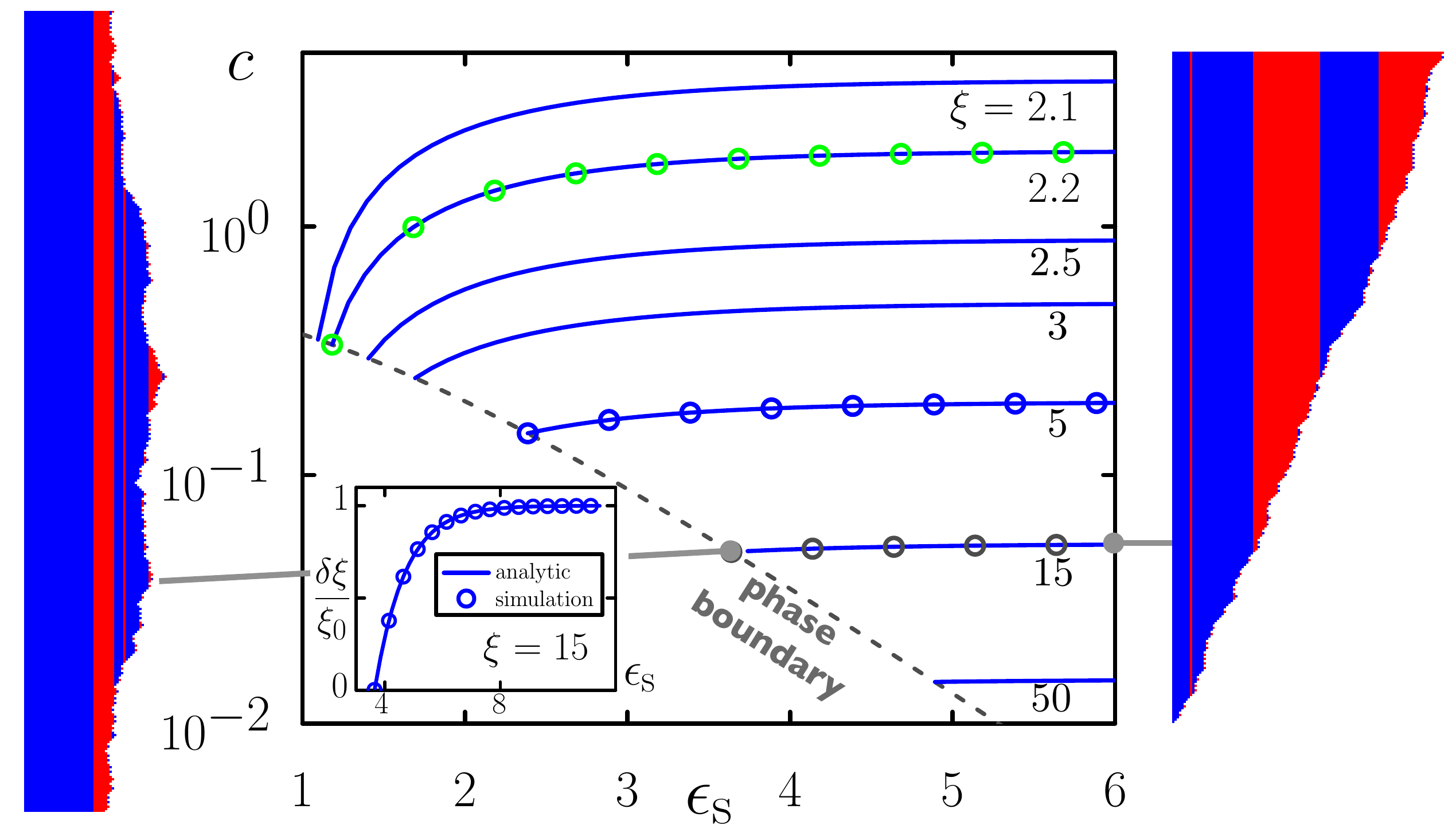}
\caption{\label{fig2} {\em Defined structures can be grown in equilibrium and far from it}. Along each contour, defined by \eqq{eq4}, identical structures can be generated by `reversible' dynamic protocols (at the phase boundary), or nearly irreversible ones (far from the phase boundary), illustrated by the simulation snapshots. Inset: `distance' $(\xi_0-\xi)/\xi_0$ to equilibrium of structures generated along the $\xi=15$ contour.}
\end{figure} 

Comparison of simulation and analytic theory thus reveals that a
fiber's dynamically-generated structure can be understood by
considering only the relative {\em net} rates of bulk domain
generation and fiber elongation. Although the time a given site has in
which to relax is governed by the number of times the fiber end
diffuses back and forth across it, with sites far from the fiber end
being `locked in'~\cite{scarlett2011mechanistic}, the structures
generated by our fluctuating simulation protocol are explained solely
by the competition between the compositional statistics of equilibrium
and random mixing, with the departure from quasiequilibrium following
from this competition enacted at a finite growth rate. The rate
of this departure, given to lowest order by \eqq{eq_pert}, has
practical consequences for the design of fiber patterns. For target
structures mimicking conventional binary crystals, having $\xi_0
\approx 1$, the conventional `perturbative' scheme of growing some
small distance $\delta c$ from the phase boundary leads to a structure
with few errors, whose compositional correlations are numerically
close to the target one (\href{http://nanotheory.lbl.gov/people/fiber_paper/fiber_supp.pdf}{Fig. S3}). We note that simulated 3d binary crystals can be
grown close to compositional
equilibrium~\cite{scarlett2011mechanistic}. However, when the desired
target possesses long range compositional correlations, growth even
close to the phase boundary results in many errors (\href{http://nanotheory.lbl.gov/people/fiber_paper/fiber_supp.pdf}{Fig. S4}). Similar
predominance of kinetics is seen in weakly-driven growth of segregated
binary assemblies~\cite{kim2008probing,scarlett2010computational}.

The `perturbative' method of self-assembly thus becomes increasingly unattractive as the compositional correlations of the target structure become increasingly complex. Here we propose an alternative `nonperturbative' approach. Our model's dynamics, \eqq{eq3}, can be cast in the scaling form
 \beq 
 \label{eq_scale}
 c/c_0 = g(\xi_0)/g(\xi),
 \eeq 
 where $g(\xi) \equiv \xi (\xi-2)/(\xi-1)$ is a scaled correlation length. This expression reveals that the results of dynamic simulations can be collapsed onto a straight line connecting a manifold of nonequilibrium conditions with an equilibrium point at $(1,1)$. We verified this collapse for a range of simulations carried out at 40 different phase points (\f{fig1}(c)) whose values of $\es$ ranged from 2 to 6. The nature of the master curve demonstrates that no fiber having $\xi = \xi_0$ can be generated anywhere past the phase boundary $c=c_0$, and emphasizes the sensitivity of structure $\xi$ to preparation condition $c$. But it also describes a useful connection between patterns in and out of equilibrium. It can be rearranged to read 
\beq
\label{eq4}
c=\frac{2(\xi-1)}{\xi (\xi-2)}\left( \emed-\emes \right), 
\eeq 
revealing that structures of identical $\xi$ can be generated on a manifold of parameters $c$, $\es$ and $\ed$, regardless of distance to the phase boundary. We illustrate this manifold in \f{fig2}. $\xi$ is constant along the displayed contour lines. Simulations verify this prediction (\href{http://nanotheory.lbl.gov/people/fiber_paper/fiber_supp.pdf}{Fig. S5}): structures generated dynamically at the circles on each contour line are indistinguishable. However, fiber generation changes from being purely diffusive (and very slow) at the phase boundary, to being nearly irreversible (and very fast) far from it, illustrated by the space-time trajectories shown (\href{http://nanotheory.lbl.gov/people/fiber_paper/fiber_supp.pdf}{Fig. S5}). Moving rightwards along contours, the generated structures lie increasingly far from the equilibrium one for the corresponding energetic parameters (\f{fig2} inset); nonetheless, structures characteristic of equilibrium at a particular point on the phase boundary can be generated by a continuum of nonequilibrium protocols away from it. In this sense, equilibrium structures can be grown, error-free, arbitrarily far from equilibrium.

{\em Conclusions.} Compositionally inhomogeneous structures are found
abundantly in biology~\cite{Brown1998membranes,Takahashi2002amyloids}
and are increasingly the target of designed
self-assembly~\cite{Kato2002liquidcrystals,Stupp2005coassembly,Glotzer2007blocks,Schulman2009AlgAssembly,kim2008probing,velikov2002layer}. We
have shown that the growth of a two-component fiber whose
compositional correlations are not random never results in the
equilibrium structure, recapitulating the importance of kinetic
factors to the self-assembly of multicomponent structures identified
in previous
theoretical~\cite{kremer1978multi,stauffer1976kinetic,PhysRevB.27.7372,schmelzer2004nucleation,schmelzer2000reconciling},
simulation~\cite{sanz2007evidence,kim2008probing,scarlett2010computational,scarlett2011mechanistic}
and experimental~\cite{kim2008probing} work. It is increasingly clear that for
multicomponent self-assembly, specification of a free energy
surface~\cite{Glotzer2004patchy,Brenner2011components} is in general not enough to ensure assembly of the
desired structure, even under conditions of weak driving. Instead, explicit accounting of how microscopic
dynamics~\cite{SchulmanWinfree2009} select assembly
pathways~\cite{Gracias2011polyhedra,Glotzer2012pathways} is
required. Indeed, we have shown that by exploiting dynamic scaling
of the rates of basic microscopic processes, precisely-defined structures can be
self-assembled out of equilibrium. Two questions arise
naturally from this work: 1) Which classes of real
multicomponent materials can be grown in quasiequilibrum? And 2) for
those that cannot, is the `nonperturbative' self-assembly strategy
described here viable? We speculate that the answer to question 1) is
`those structures displaying local free energy barriers on the scale
of the desired compositional correlation length'. 3d binary crystals with different
lattice structures appear to display different propensities for
growing in compositional equilibrium~\cite{scarlett2011mechanistic},
and we speculate that the free energy cost of adding a few particles
to an existing structure may be a key difference in this
regard. However, when the desired compositional correlation length is
large, achieving free energy barriers over the extent of this length is likely unworkable. We then turn to question
2), which experimental evidence hints might be answered in the
positive. The dynamic scaling that permits the far-from-equilibrium
assembly strategy we have described is strikingly reminiscent of data
collapse seen in segregated binary structure growth, where an
effective structural order parameter scales with effective crystal
growth velocity~\cite{kim2008probing}. Such scaling, occurring in a
regime in which crystals are morphologically ordered, suggests the
possibility of designing compositionally ordered structures in
equilibrium, and assembling them far from it.

{\em Acknowledgements.} We thank Tom Haxton, Rob Jack, and David Sivak for discussions. SW was supported by the Director, Office of Science, Office of Basic Energy Sciences, of the U.S. Department of Energy under Contract No. DE-AC02--05CH11231. R.S was supported by Johns Hopkins. L.O.H. was supported by the Center for Nanoscale Control of Geologic CO$_2$, a U.S. D.O.E. Energy Frontier Research Center, under Contract No. DE-AC02--05CH11231.


\end{document}